# Social Content Matching in MapReduce

Gianmarco De Francisci Morales[*]
IMT Lucca and ISTI-CNR Pisa
Italy
gdfm@yahoo-inc.com

Aristides Gionis
Yahoo! Research
Barcelona, Spain
gionis@yahoo-inc.com

Mauro Sozio[†]
Max-Planck-Institut für Informatik
Saarbrücken, Germany
msozio@mpi-inf.mpg.de

## ABSTRACT

Matching problems are ubiquitous. They occur in economic markets, labor markets, internet advertising, and elsewhere. In this paper we focus on an application of matching for social media. Our goal is to distribute content from information suppliers to information consumers. We seek to maximize the overall relevance of the matched content from suppliers to consumers while regulating the overall activity, e.g., ensuring that no consumer is overwhelmed with data and that all suppliers have chances to deliver their content.

We propose two matching algorithms, GREEDYMR and STACKMR, geared for the MapReduce paradigm. Both algorithms have provable approximation guarantees, and in practice they produce high-quality solutions. While both algorithms scale extremely well, we can show that STACK-MR requires only a poly-logarithmic number of MapReduce steps, making it an attractive option for applications with very large datasets. We experimentally show the trade-offs between quality and efficiency of our solutions on two large datasets coming from real-world social-media web sites.

## 1. INTRODUCTION

The last decade has witnessed a fundamental paradigm shift on how information content is distributed among people. Traditionally, the majority of information content has been produced by few specialized agents and consumed by the big masses. Nowadays, an increasing number of platforms allow everyone to participate both in information production and in information consumption. The phenomenon has been coined as *democratization of content*. The Internet, and its younger children, user-generated content and social media, have had a major role in this paradigm shift.

Blogs, micro-blogs, photo-sharing systems, and question-answering systems, are some of the social media that people participate in as both information suppliers and information consumers. In such social systems, not only consumers have many opportunities to find relevant content, but also suppliers have opportunities to find the right audience for their content. However, as the opportunities to find relevant information and relevant audience increase, so does the complexity of a system that would allow suppliers and consumers to meet in the most efficient way.

Our motivation is building a *"featured item"* component for social-media applications. Such a component would provide recommendations to consumers each time they log in the system. For example, `flickr` displays photos to users when they enter their personal pages, while `Yahoo! Answers` displays questions that are still open for answering. For consumers it is desirable that the recommendations are of high quality and relevant to their interests. For suppliers it is desirable that their content is delivered to consumers who are interested in it and may provide useful feedback. In this way, both consumers and suppliers are more satisfied by using the system and they get the best out of it.

Naturally, we model this problem as a matching problem. We associate a *relevance* score to each potential match of an item $t$ to a user $u$. This score can be seen as the weight of the edge $(t, u)$ of the bipartite graph between items and users. For each item $t$ and each user $u$ we also consider constraints on the maximum number of edges that $t$ and $u$ can participate in the matching. These *capacity constraints* can be estimated by the activity of each user and the relative frequency with which items need to be delivered. The goal is to find a matching that satisfies all capacity constraints and maximizes the total weight of the edges in the matching. This problem is known as *b-matching*.

The *b*-matching problem can be solved in polynomial time by max-flow techniques. However, the fastest exact algorithms today have complexity $\tilde{O}(nm)$ [10, 13], for graphs with $n$ nodes and $m$ edges, and thus do not scale to large datasets. Instead in this paper we focus on approximation algorithms that are scalable to very large datasets. We propose two *b*-matching algorithms, STACKMR and GREEDY-MR, which can be implemented efficiently in the *MapReduce* paradigm [8]. While both our algorithms have provable approximation guarantees, they have different properties.

We design the STACKMR algorithm by drawing inspiration from existing distributed algorithms for matching problems [12, 21]. STACKMR is allowed to violate capacity constraints by a factor of $(1 + \epsilon)$, and yields an approximation

---

guarantee of $\frac{1}{6+\epsilon}$, for any $\epsilon > 0$. We show that STACKMR requires a poly-logarithmic number of MapReduce steps. This makes STACKMR appropriate for realistic scenarios with large datasets. We also study a variant of STACKMR, called STACKGREEDYMR, in which we incorporate a greedy heuristic in order to obtain higher-quality results.

On the other hand, GREEDYMR is a simpler algorithm to implement and has the desirable property that it can be stopped at any time and provide the current best solution. GREEDYMR has a better quality guarantee than STACKMR, as it is a $\frac{1}{2}$-approximation algorithm. GREEDYMR also yields better solutions in practice. However, it cannot guarantee a poly-logarithmic number of steps. A simple example shows that GREEDYMR may require a linear number of steps. Although GREEDYMR is theoretically less attractive than STACKMR, in practice it is a very efficient algorithm, and its performance is way far from the worst case.

Finally, we note that the $b$-matching algorithm takes as input the set of candidate edges weighted by their relevance scores. In some cases, this set of candidate edges is small, for instance when items are recommended only among friends in a social network. In other applications, any item can be delivered to any user, e.g., a user in flickr may view a photo of any other user. In the latter case, materializing all pairs of item-user edges is an unfeasible task. Thus, we equip our framework with a scheme that finds all edges with score greater than some threshold $\sigma$, and we restrict the matching to those edges. Interestingly, finding all similar item-user pairs can also be implemented efficiently in MapReduce, by modifying recent algorithms developed for computing the self-join of a document collection [2].

Our main contributions are the following:

- We investigate the problem of $b$-matching in the context of social content distribution, and devise a fully-MapReduce framework to address it.

- We develop STACKMR, an efficient variant of the algorithm presented in [21]. We show how to adapt such an algorithm in MapReduce, while requiring only a poly-logarithmic number of steps. Our experiments show that STACKMR scales excellently to very large datasets.

- We introduce GREEDYMR, a MapReduce adaptation of a classical greedy algorithm. It has a $\frac{1}{2}$-approximation guarantee, and is very efficient in practice.

- We employ recent techniques for similarity self-join in MapReduce to build the input graph for the $b$-matching.

- We perform a thorough experimental evaluation using large datasets extracted from real-world scenarios.

## 2. RELATED WORK

The general problem of assigning entities to users so to satisfy some constraints on the overall assignment arises in many different research areas of computer science. Entities could be advertisements [3], items in an auction [22], scientific papers [11] or media content, like in our case. The $b$-matching problem finds applications also in machine learning [15] and in particular in spectral clustering [14].

The weighted $b$-matching problem can be solved in polynomial time by employing maximum flow techniques [10, 13], however, the running time is still superlinear in the worst case. A faster approximation algorithm has been recently proposed by Christiano et al. [5].

In a distributed environment, there are some results for the unweighted version of the (simple) matching problem [9, 12], while for the weighted case the approximation guarantee has progressively improved from $\frac{1}{5}$ [24] to $\frac{1}{2}$ [20]. For distributed weighted $b$-matching, a $\frac{1}{2}$-approximation algorithm was developed by Koufogiannakis and Young [18]. However, a MapReduce implementation is non-obvious.

The MapReduce paradigm [8] has been designed to deal with the huge amount of data that is readily available nowadays. Many existing algorithms in data mining and machine learning have been adapted to MapReduce [4, 6, 19]. Karloff et al. [17] give an abstract model of computation for MapReduce. Chierichetti et al. [4] and Karloff et al. [17] develop algorithms for the classic problems of max cover and minimum-spanning tree, respectively. However, MapReduce implementations of algorithms for graph problems have been relatively limited so far [7, 16]. A classical model for parallel computation is the PRAM model, where processors have access to a shared memory which can also be used to communicate. It has been shown that many algorithms developed for the PRAM model can be adapted in MapReduce [17].

## 3. PRELIMINARIES

### 3.1 The MapReduce model

**MapReduce** [8] is a distributed computing paradigm based on two higher-order functions: map and reduce. The map function applies a user-defined function to each key-value pair in the input. The result is a list of intermediate key-value pairs, sorted and grouped by key, and passed as input to the reduce function. The reduce function applies a second user-defined function to every intermediate key and all its associated values, and produces the final result. The signatures of the functions that compose the phases of a MapReduce computation are as follows:

$$\text{map}: \quad \langle k_1, v_1 \rangle \quad \rightarrow \quad [\langle k_2, v_2 \rangle]$$
$$\text{reduce}: \quad \langle k_2, [v_2] \rangle \quad \rightarrow \quad [\langle k_3, v_3 \rangle]$$

MapReduce assumes a distributed file system from which the map instances retrieve the input. The framework takes care of moving, grouping, and sorting the intermediate data. This phase is called *shuffle*, and strongly affects the efficiency of any MapReduce-based implementation. In our work we used hadoop, an open-source implementation of MapReduce.

### 3.2 Problem definition

In this section we introduce our notation and provide our problem formulation. We are given a set of content items $T = \{t_1, \ldots, t_n\}$, which are to be delivered to a set of consumers $C = \{c_1, \ldots, c_m\}$. For each $t_i$ and $c_j$, we assume we are able to measure the interest of consumer $c_j$ in item $t_i$ with a positive weight $w(t_i, c_j)$. The distribution of the items $T$ to the consumers $C$ can be clearly seen as a matching problem on the bipartite graph with nodes $T$ and $C$, and edge weights $w(t_i, c_j)$.

In order to avoid that each consumer $c_j$ receive too many items, we enforce a capacity constraint $b(c_j)$ on the number of items that are matched to $c_j$. Similarly, we would like to avoid the scenario when only a few items (e.g. the most popular ones) participate in the matching. To this end, we introduce a capacity constraint $b(t_i)$ on the number of consumers that each item $t_i$ is matched to.

This variant of the matching problem is well known in the theoretical computer science community as the $b$−matching problem. This is defined as follows. We are given an undirected graph $G = (V, E)$ and a function $b : V \to \mathbb{N}$ expressing node capacities (or budgets). Every edge $e$ comes with a positive weight $w(e)$. A $b$-matching in $G$ is a subset of $E$ such that for each node $v$ in $V$ at most $b(v)$ edges incident to $v$ are in the matching. We wish to find a $b$-matching of maximum weight.

Although all our algorithms can deal with any undirected graph, we focus on bipartite graphs which are relevant in our application scenarios. The problem we shall consider in the rest of the paper is then defined as follows.

**Problem 1** *We are given an undirected bipartite graph $G = (T, C, E)$, where $T$ represents a set of items and $C$ represents a set of consumers, a weight function $w : E \to \mathbb{R}^{+}$, as well as a capacity function $b : T \cup C \to \mathbb{N}$. A $b$-matching in $G$ is a subset of $E$ such that for each node $v$ in $T \cup C$ at most $b(v)$ edges incident to $v$ are in the matching. We wish to find a $b$-matching of maximum weight.*

## 4. APPLICATION SCENARIOS

To instantiate the problem we just defined, we need to $(i)$ define the weights $w(t_i, c_j)$ between items $t_i$ and consumers $c_j$, $(ii)$ decide the set of potential edges that participates in the matching, and $(iii)$ define the capacity constraints $b(t_i)$ and $b(c_j)$. In this paper we focus only on the matching algorithm and we assume that addressing the details of the above questions depends on the application. However, for completeness we discuss our thoughts on the above issues.

**Scenario.** We envision a scenario in which an application operates in consecutive phases. The duration of each phase may range from hours to days. Before the beginning of the $i$-th phase the application makes a tentative allocation of which items will be delivered to which consumers during the $i$-th phase. The items that participate in this allocation, i.e., the set $T$ of Problem 1, are those that have been produced during the $(i-1)$-th phase, and perhaps other items that have not been distributed in previous phases.

**Edge weights.** A simple approach is to represent items and consumers in a *vector space*, i.e., items $t_i$ and consumers $c_j$ are represented by *term vectors* $\mathbf{v}(t_i)$ and $\mathbf{v}(c_j)$. Then we can define the edge weight $w(t_i, c_j)$ using the dot-product similarity $w(t_i, c_j) = \mathbf{v}(t_i) \cdot \mathbf{v}(c_j)$. More complex similarity functions can be used, too. Borrowing ideas from information retrieval, the terms in the vector representation can be weighted with `tf·idf` scores. Alternatively, the weights $w(t_i, c_j)$ could be the output of a *recommendation system* that takes into account user preferences and user activities.

**Candidate edges.** With respect to deciding which edges to consider for matching, the simplest approach is to consider all possible pairs $(t_i, c_j)$. This is particularly attractive, since we let the decision of selecting edges entirely to the matching algorithm. However, considering $O(|T||C|)$ edges makes the system highly inefficient. Thus, we opt for methods that prune the number of candidate edges. Our approach is to consider as candidates only edges whose weight $w(t_i, c_j)$ is above a threshold $\sigma$. The rationale is that since the matching algorithm will seek to maximize the total edge weight, we preferably discard low-weight edges.

We note that depending on the application, there may be other ways to define the set of candidate edges. For example, in social-networking sites it is common for consumers to *subscribe* to suppliers they are interested in. In such an application, we restrict to candidate edges $(t_i, c_j)$ for which $t_i$ has been created by a producer to whom $c_j$ has subscribed.

**Capacity constraints.** The consumer capacity constraints express the number of items that need to be displayed to each consumer. For example, if we display one different item to a consumer each time they access the application, $b(c_j)$ can be set to an estimate of the number of times that consumer $c_j$ will access the application during the $i$-th phase. Such an estimate can be obtained from log data.

For the item capacity constraints, we observe that $B = \sum_{c \in C} b(c)$ is an upper bound on the total number of distributed items, so we require $B = \sum_{t \in T} b(t)$ as well. Now we distinguish two cases, depending on whether there is a *quality assessment* on the items $T$ or not. If there is no quality assessment, all items are considered equivalent, and the total distribution bandwidth $B$ can be divided equally among all items, so $b(t) = \max\{1, \frac{B}{|T|}\}$, for all $t$ in $T$.

If there is a quality assessment on the items $T$, we assume a quality estimate $q(t)$ for each item $t$. Such an estimate can be computed using a machine-learning approach, as the one proposed by Agichtein et al. [1], which involves various features like content, links, and reputation. Without loss of generality we assume normalized scores, i.e., $\sum_{t \in T} q(t) = 1$. We can then divide the total distribution bandwidth $B$ among all items in proportion to their quality score, so $b(t) = \max\{1, q(t)B\}$. In a real-application scenario, the designers of the application may want to control the function $q(t)$ so that it satisfies certain properties, for instance, it follows a power-law distribution.

## 5. ALGORITHMS

### 5.1 Computing the set of candidate edges

The first step of our algorithm is to compute the set of candidate edges, which in Section 4 were defined to be the edges with weight $w(t_i, c_j)$ above a threshold $\sigma$. This step is crucial in order to avoid considering $O(|T||C|)$ edges, which would make the algorithm impractical.

The problem of finding all the pairs of $t_i \in T$ and $c_j \in C$ so that $w(t_i, c_j) \geq \sigma$ is known as the *similarity join* problem. Since we aim at developing the complete system in the MapReduce framework, we take advantage of recent advances in the problem of computing the *similarity self-join between documents* in MapReduce. In particular, we adapt the algorithm of Baraglia et al. [2], which to our knowledge is the state-of-the-art in computing self-join in MapReduce.

The algorithm of Baraglia et al. works using the technique of *prefix filtering*: the main idea is to create a *pruned inverted index* on the whole set of documents and then query those documents on the index. The pruning of the index guarantees that for documents $d_j$ not retrieved during the querying by document $d_i$ the similarity between $d_i$ and $d_j$ is below the minimum threshold. For each pair of documents returned in the querying phase the algorithm needs to access the documents and verify if the similarity is indeed greater than the threshold. The algorithm of Baraglia et al. implements all the above tasks efficiently in MapReduce. Overall two MapReduce iterations are required.

Our algorithm for computing the set of candidate edges is a simple modification of the algorithm of Baraglia et al. First we interpret the items $t_i$ and the consumers $c_j$ as documents using their vector representation. Second the self-join algorithm can be modified to join the two sets $T$ and $C$ without considering pairs between two items or two consumers.

## 5.2 The stack algorithm

Our first matching algorithm, STACKMR, is a variant of the algorithm developed in [21]. The main difference is that in [21] there is an involved mechanism to ensure that capacity constraints are satisfied, which unfortunately does not seem to have an efficient implementation in MapReduce. Here we devise a more practical variant that allows node capacities to be violated by a factor of at most $(1 + \epsilon)$, for any $\epsilon > 0$. This is a small price to pay in our application scenarios, where small capacity violations can be tolerated.

For the sake of presentation, we describe our algorithm first in a centralized environment and then in a parallel environment. Pseudocode for STACKMR and for the algorithm in [21] are included in the Appendix. (Algorithm 2 and 1, respectively). The latter one has been slightly changed so to take into account implementation issues. However, we do not include an evaluation of this algorithm as it does not seem to be efficient. In the next section we describe in detail how to implement the former algorithm in MapReduce.

Our algorithm is based on the *primal-dual schema*, a successful and established technique to develop approximation algorithms. The primal-dual schema has proved to play an important role in the design of sequential and distributed approximation algorithms ([21, 23]). We believe that primal-dual algorithms bear the potential of playing an important role in the design of MapReduce algorithms as well.

The first step of any primal-dual algorithm is to formulate the problem at hand as an integer linear program (IP). Consequently, integrality constraints are relaxed so that variables can take any value in the range $[0, 1]$. This linear program (LP) is called *primal*. From the primal program we can derive the so-called *dual*. There is a direct correspondence between the variables of the primal and the constraints of the dual, as well as, the variables of the dual and the constraints of the primal.

A typical primal-dual algorithm proceeds as follows: at each step dual variables are raised and as soon as a dual constraint is satisfied with equality (or almost) the corresponding primal variable is set to one. The above procedure is iterated until when no dual variable can be increased further without violating a dual constraint. The approximation guarantee follows from the fact that any feasible dual solution gives an upper bound on any optimum solution for the primal. Thus, the closer these two quantities be the better the approximation guarantee. For a more exhaustive illustration of the technique, see the book of Vazirani [23].

We now present our algorithm in a centralized environment. We first give the IP for the *b*-matching problem.

$$\text{maximize} \quad \sum_{e \in E} w(e) x_e \qquad \text{(IP)}$$

$$\text{such that} \quad \sum_{e \in E, v \in e} x_e \leq b(v) \qquad \forall v \in V, \quad (1)$$

where $x_e \in \{0, 1\}$ is associated to edge $e$, and a value of 1 means that $e$ belongs to the solution. Equation (1) expresses capacity constraints. The dual program is as follows.

$$\text{minimize} \quad \sum_{v \in V} y_v \qquad \text{(DP)}$$

$$\text{such that} \quad \frac{y_u}{b(u)} + \frac{y_v}{b(v)} \geq w(e) \quad \forall e = (u, v) \in E, \quad (2)$$

$$y_v \geq 0 \qquad \forall v \in V. \quad (3)$$

Dual constraints (2) are associated with edges $e$. An edge is said to be *covered* if its corresponding constraint is satisfied with equality. The variables occurring in such a constraint are referred as $e$'s dual variables and play an important role in the execution of the algorithm.

The centralized algorithm consists of two phases: a *push phase* where edges are pushed on a stack in arbitrary order, and a *pop phase* where edges are popped from the stack and a feasible solution is computed. When an edge $e(u, v)$ is pushed on the stack, each of its dual variables is increased by the same amount $\delta(e)$ so to satisfy Equation (2) with equality. The amount $\delta(e)$ is derived from Equation (2) as

$$\delta(e) = \frac{(w(e) - y_u/b(u) - y_v/b(v))}{2}. \quad (4)$$

Whenever edges become covered they are deleted from the input graph. The push phase terminates when no edge is left. In the pop phase, edges are successively popped out of the stack and included in the solution if feasibility is maintained. This discussion concludes the description of the centralized algorithm.

In a parallel environment, we wish to parallelize as many operations as possible so to ensure poly-logarithmic running time. Thus, we need a mechanism to bound the number of push and pop steps, which in the centralized algorithm may be linear in the number of edges. This is done by computing at each step a *maximal [ɛb]-matching*. Note the difference between maximum and maximal: a *b*-matching is *maximal* if and only if it is not properly contained in any other *b*-matching. All edges in a maximal set, called a *layer* of the stack, are pushed on the stack in parallel. In the popping phase, all edges within the same layer are popped out of the stack and included in the solution in parallel. Edges of nodes whose capacity constraints are satisfied or violated are deleted from the stack and ignored from further consideration. A maximal *b*-matching can be computed efficiently in MapReduce as we will discuss in Section 5.3.

Unfortunately, the total number of layers may still be linear in the maximum degree of a node. To circumvent this problem, we introduce the definition of *weakly covered edges*. Roughly speaking, a weakly covered edge is an edge whose constraint is only "partially satisfied" and thus gets covered after a few number of iterations. A formal definition follows.

**Definition 1 (Weakly covered edges)** *Given $\epsilon > 0$, at any time during the execution of our algorithm we say that an edge $e \in E$ is weakly covered if constraint (2) for $e = uv$ is such that*

$$\frac{\bar{y}_u}{b(u)} + \frac{\bar{y}_v}{b(v)} \geq \frac{1}{3 + 2\epsilon} w(e), \quad (5)$$

*where $\bar{y}$ denotes the current value of $y$.*

Observe that Equation (5) is derived from Equation (2).

To summarize, our parallel algorithm proceeds as follows. At each step of the push phase, we compute a maximal

$\lceil \epsilon b \rceil$-matching using the procedure by Garrido et al. [12]. All the edges in the maximal matching are then pushed on the stack in parallel forming a layer of the stack. For each of these edges we increase each of its dual variable by $\delta(e)$ in parallel. Some edges might then become weakly covered and are deleted from the input graph. The push phase is executed until no edge is left.

At the end of the push phase, layers are iteratively popped out of the stack and edges within the same layer are included in the solution in parallel. This can violate node capacities by a factor of at most $(1+\epsilon)$, as every layer contains at most $\epsilon b(v)$ edges incident to any node $v$. Edges of nodes whose capacity constraints are satisfied or violated are deleted from the stack and ignored from further consideration. This phase is iterated until the stack becomes empty.

We can show that the approximation guarantee of our algorithm is $\frac{1}{6+\epsilon}$, for every $\epsilon > 0$. Moreover, we can show that the push phase is iterated at most $O(\log \frac{w_{max}}{w_{min}})$ steps, where $w_{max}$ and $w_{min}$ are the maximum and minimum weight of any edge in input, respectively. This together with the fact that the procedure in [12] requires $O(\log^3 n)$ rounds imply the following theorem.

**Theorem 1** *Algorithm 2 has an approximation guarantee of $\frac{1}{6+\epsilon}$ and violates capacity constraints by a factor of at most $1 + \epsilon$. It requires $O(\frac{\log^3 n}{\epsilon^2} \cdot \log \frac{w_{max}}{w_{min}})$ communication rounds, with high probability.*

The non-determinism of the algorithm follows from the non-determinism of the algorithm that computes maximal $b$-matchings. The proof of Theorem 1 is very similar to the proof in [21] and is omitted for lack of space. STACKMR is a factor of $\frac{1}{\epsilon}$ faster than the the algorithm presented in [21].

### 5.3 Adaptation in MapReduce

The distributed algorithm described in the previous section works in an iterative fashion. In each iteration we first compute a *maximal matching*, then we *push* it in a stack, we *update* edges, and we *pop* all levels from the stack. Below we describe how to implement these steps in MapReduce.

**Maximal matching.** For finding maximal $b$-matchings we employ the algorithm of Garrido et al. [12], which is an iterative probabilistic algorithm. Each iteration consists of four stages: (*i*) `marking`, (*ii*) `selection`, (*iii*) `matching`, and (*iv*) `cleanup`. The stages are as follows.

In the `marking` stage, each node $v$ *marks* randomly $\lceil \frac{1}{2}b(v) \rceil$ of its incident edges. In the `selection` stage, each node $v$ *selects* randomly $\max\{\lfloor \frac{1}{2}b(v) \rfloor, 1\}$ edges marked by its neighbors. At this point, a set of edges $F$ have been selected. In the `matching` stage, if some node $v$ has capacity $b(v) = 1$ and two incident edges in $F$, it randomly deletes one of them. At this point the set $F$ is a valid $b$-matching. The set $F$ is added to the solution and removed from the original graph. In the `cleanup` stage, each node updates its capacity in order to take into consideration the edges in $F$ and saturated nodes are removed from the graph. These stages are iterated until there are no more edges left in the original graph. Garrido et al. [12] show that the process requires, on expectation, $O(\log^3 n)$ iterations to terminate.

To adapt this algorithm in MapReduce, we need one job for each stage. The input and output of each MapReduce job is always of the same format: a consistent view of the graph represented as adjacency lists. We maintain a "node-based" representation of the graph because we need to make decisions based on the local neighborhood of each node. Assuming the set of nodes adjacent to $v$ is $\{v_1, \ldots, v_k\}$, the input and output of each job is a list $\langle v_i, [(v_j, T_j), \ldots, (v_k, T_k)] \rangle$, where $v_i$ is the key and $[(v_j, T_j), \ldots, (v_k, T_k)]$ the associated value. The variables $T$ represent the *state* of each edge. We consider five possible states of an edge: $E$ in the main graph; $K$ marked; $F$ selected; $D$ deleted; and $M$ in the matching.

Each `map` function performs the decisions altering the state of the graph locally to each node. Each `reduce` function unifies the diverging views of the graph at each node. For each edge $(v_i, v_j)$, each `map` function will emit both $\langle v_i \rangle$ and $\langle v_j \rangle$ as keys, together with the current state of the edge as value. The `reduce` function will receive the views of the state of the edge from both end-points, and will unify them, yielding a consistent graph representation as output.

Each MapReduce job uses the same communication pattern and state unification rules. They only differ in the way they update the state of the edges. The communication cost of each job is thus $O(|E|)$, while the achievable degree of parallelism is $O(|V|)$.

**Push, update, and pop.** The basic scheme of communication for the *push*, *update*, and *pop* phases is the same as the one for computing maximal matching. For these phases of the algorithm, we maintain a separate state for each edge. The possible states in which an edge can be are: $E$, edge in the graph (default); $S$, edge stacked; $R$, edge removed from the graph; and $I$, edge included in the solution. For each edge we also maintain an integer variable that represents the stack level in which the edge has been put.

During the push phase, for each edge included in the maximal matching, we set its state to $S$ and the corresponding stack level. The update phase is needed to propagate the $\delta(e)$ contributions. Each edge sent to a node $v$ carries the value of its sending node $y_u/b(u)$. Thus, each node can compute the new $\delta(e)$ and update its local $y_v$. This phase also removes weakly covered edges by setting their state to $R$, and updates the capacities of the nodes for the next maximal-matching phase. Removed edges are not considered for the next maximal-matching phase. When all the edges in the graph are either stacked ($S$) or removed ($R$), the pop phase starts. During the pop phase, each stacked ($S$) edge in the current level (starting from the topmost) is included in the solution by setting its state to ($I$). The capacities are locally updated, and nodes (and all incident edges) are removed when their capacity becomes non-positive.

### 5.4 The greedy algorithm

In this section we present a second matching algorithm based on a greedy strategy: GREEDYMR. We analyze the centralized version and then we adapt it in MapReduce.

The centralized greedy algorithm works by processing sequentially each edge in order of decreasing weight. It includes an edge $e(u, v)$ in the solution if $b(u) > 0$ and $b(v) > 0$. In this case, it subtracts 1 from both $b(u)$ and $b(v)$.

It is immediate that the greedy algorithm produces a feasible solution. In addition, it has a factor $\frac{1}{2}$ approximation guarantee. We believe that this is a well-known result, however, we were not able find a reference. Thus, for completeness we include a proof in the Appendix.

The GREEDYMR algorithm is a MapReduce adaptation of the above centralized algorithm. We note that the adapta-tion is not straightforward, due to the access to the globally-shared variables $b(v)$ that keep node capacities.

GREEDYMR works as follows. In the `map` phase each node $v$ proposes its $b(v)$ edges with maximum weight to its neighbors. In the `reduce` phase, each node computes the intersection between its own proposals and the proposals of its neighbors. The set of edges in the intersection is included in the solution. Then, each node updates its capacity. If it becomes 0, the node is removed from the graph. A pseudocode for GREEDYMR is shown in the Appendix (Algorithm 3).

In contrast with STACKMR, GREEDYMR is not guaranteed to finished in a poly-logarithmic number of iterations. As a simple worst-case input instance, consider a path graph $u_1u_2, u_2u_3, ...u_{k-1}u_k$ such that $w(u_i, u_{i+1}) \leq w(u_{i+1}, u_{i+2})$. GREEDYMR will face a chain of cascading updates that will cause a linear number of MapReduce iterations. However, as we will see in Section 6, in practice GREEDYMR yields quite competitive results compared to STACKMR.

Finally, an additional advantage of GREEDYMR is that it maintains a feasible solution at each step. Therefore the algorithm can be terminated at any step and return the current solution. This property makes the algorithms especially attractive in our application scenarios, where content can be delivered to the users almost immediately and the algorithm can continue running in the background.

# 6. EXPERIMENTAL EVALUATION

In this section we describe the experiments we perform. We start with the description of the datasets we use.

**Flickr.** We extract two datasets from `flickr`, a photo-sharing site. Table 1 shows statistics for the `flickr-small` and `flickr-large` datasets. In these datasets items represent photos and consumers represent users.

Recall from our discussion in Section 4 that the capacity $b(u)$ of each user $u$ should be set in proportion to the login activity of the user in the system. Unfortunately, the login activity is not available in the dataset, so we decide to use as a *proxy* the number of photos $n(u)$ that the user $u$ has posted. We then use a parameter $\alpha > 0$ to set the capacity of each user $u$ as $b(u) = \alpha\,n(u)$. Higher values of the parameter $\alpha$ simulate higher levels of activity in the system.

Next we need to specify the capacity of photos. Since our primary goal is to study the matching algorithm, specifying the actual capacities is beyond the scope of the paper. Thus we use as a *proxy*, the number of favorites $f(p)$ that each photo $p$ has received. The intuition is that we want to favor good photos in order to increase user satisfaction. Following Section 4, we set the capacity of each photo to

$$b(p) = f(p)\frac{\sum_u \alpha n(u)}{\sum_q f(q)}.$$

In order to estimate edge similarities, we represent each photo by its tags, and each user by the set of all tags he or she has used. Then we compute the similarity between a photo and a user as the dot product of the their tag vectors. We compute all edges whose similarity is larger than a threshold $\sigma$ using the MapReduce algorithm described in Section 5.1. The distributions of edge similarities are shown in Figure 6 in the Appendix.

**Yahoo! Answers.** We extract one dataset from the `Yahoo! Answers` question-answering portal. In `yahoo-answers`, consumers represent users, while items represent questions. The motivating application is to propose unanswered questions to users. Matched questions should fit the user interests. To identify user interests, we represent users by the weighted set of words in their answers. We preprocess the answers to remove punctuation and stop-words, stem words, and apply `tf·idf` weighting. We treat questions similarly.

**Table 1: Dataset characteristics.** $|T|$: number of items; $|C|$: number of users; $|E|$: total number of item-user pairs with non zero similarity.

| Dataset | $|T|$ | $|C|$ | $|E|$ |
|---|---|---|---|
| `flickr-small` | 2 817 | 526 | 550 667 |
| `flickr-large` | 373 373 | 32 707 | 1 995 123 827 |
| `yahoo-answers` | 4 852 689 | 1 149 714 | 18 847 281 236 |

As before, we extract a bipartite graph with edge weights representing the similarity between questions and users. We employ again a threshold $\sigma$ to sparsify the graph, and present results for different density levels. In this case, we set user capacities $b(u)$ by employing the number of answers $n(u)$ provided by each user $u$ as a proxy to the activity of the user. We use the same parameter $\alpha$ as for the `flickr` datasets to set $b(u) = \alpha\,n(u)$. However, for this dataset we use a constant capacity for all questions, in order to test our algorithm under different settings. For each question $q$ we set

$$b(q) = \frac{\sum_u \alpha n(u)}{|Q|}.$$

**Variants.** We also experiment with a number of variants of the STACKMR algorithm. In particular, we vary the edge-selection strategy employed in the first phase of the maximal $b$-matching algorithm (`marking`) [12]. The STACKMR algorithm proposes to its neighbors edges chosen uniformly at random. In order to favor heavier edges in the matching, we modify the selection strategy to propose the $\lceil \frac{1}{2}\epsilon b(v) \rceil$ edges with the largest weight. We call this variant STACKGREEDY-MR. We also experiment with a third variant, in which we choose edges randomly but with probability proportional to their weights. Because it always performs worse than STACKGREEDYMR and for lack of space we do not show the results for this third variant.

**Measures.** We evaluate the proposed algorithms in terms of *quality* and *efficiency*. Quality is measured in terms of $b$-matching value achieved, and efficiency in terms of the number of MapReduce iterations required. We evaluate our algorithms by varying the following parameters: the similarity threshold $\sigma$, which controls the number of edges that participate in the matching; the slackness parameter $\epsilon$; and the factor $\alpha$ in determining capacities.

**Results.** Sample results on the quality and efficiency of our matching algorithms for the three datasets, `flickr-small`, `flickr-large`, and `yahoo-answers`, are shown in Figures 1, 2, and 3, respectively. For each plot in these figures, we fix the parameters $\epsilon$ and $\alpha$ and we vary the similarity threshold $\sigma$. Our observations are summarized as follows.

*Quality.* GREEDYMR consistently produces matchings with higher value than the two stack-based algorithms. Since GREEDYMR has better approximation guarantee, this result is in accordance with theory. In fact, GREEDYMR achieves better results even though the stack algorithms have the advantage of being allowed to exceed node capacities. However, as we will see next, the violations are very small, ranging from practically 0 to at most 6%. In the `flickr-large` dataset, GREEDYMR produces solutions that have on average 31% higher value than solutions by STACK-MR. In `flickr-small` and `yahoo-answers`, the improvement of GREEDYMR is 11% and 14%, respectively. When comparing the two stack algorithms, we see that STACK-GREEDYMR is slightly better than STACKMR. Again the difference is more pronounced on the `flickr-large` dataset.

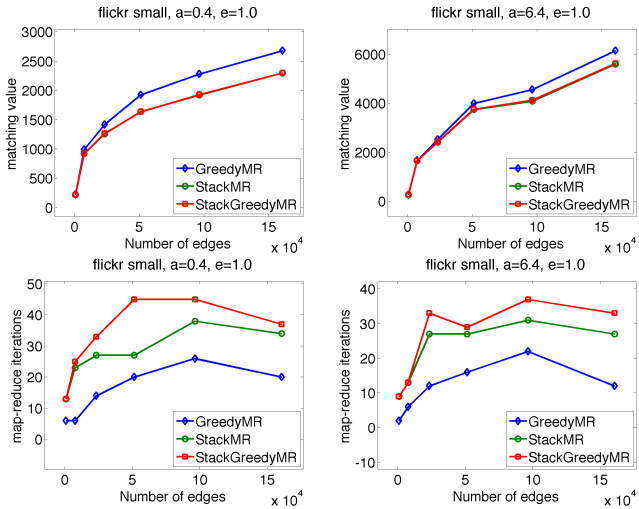

**Figure 1:** `flickr-small` dataset: matching value and number of iterations as a function of the number of edges.

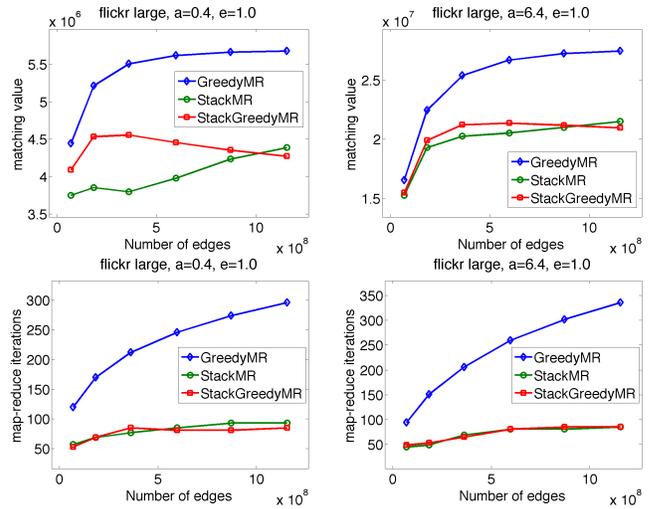

**Figure 2:** `flickr-large` dataset: matching value and number of iterations as a function of the number of edges.

We also observe that in general the $b$-matching value increases with the number edges. This behavior is expected, as the number of edges increase the algorithms have more flexibility. Since we add edges by lowering the edge-similarity threshold, the gain in the $b$-matching value tends to saturate. The only exception to this rule is for STACKGREEDYMR on the `flickr-large` dataset. We believe this is due to the uneven capacity distribution for the `flickr-large` dataset, see Figure 7. Our belief is supported by fact that the decrease is less visible for higher values of $\alpha$.

*Efficiency.* Our findings validate the theory also in terms of efficiency. In most settings the stack algorithms perform better than GREEDYMR. The only exception is the `flickr-small` dataset. This dataset is very small, so the stack algorithms incur additional overhead when computing maximal matchings. However, the power of the stack algorithms is demonstrated on the large datasets. Not only they require less MapReduce steps than GREEDYMR, but they scale extremely well. The performance of STACKMR is almost unaffected by increasing the number of edges.

*Capacity violations.* As explained in Section 5.2, STACK-MR and STACKGREEDYMR can exceed the capacity of the nodes by a factor of $(1 + \epsilon)$. However, in our experiments the algorithms exhibit much lower capacity violations than the worst case. We compute the average violation as

$$\epsilon' = \frac{1}{|V|} \sum_{v \in V} \frac{\max\{|M(v)| - b(v), 0\}}{b(v)},$$

where $|M(v)|$ is the degree of node $v$ in the matching $M$, and $b(v)$ is the capacity for each node $v$ in $V$. Figure 4 shows capacity violations for STACKMR. The violations for STACK-GREEDYMR are similar, and omitted for lack of space. For $\epsilon = 1$, for the `flickr-large` dataset the violation is as low as 6% in the worst case. As expected, more violations occur when more edges are allowed to participate in the matching, either by increasing the number of edges (lower $\sigma$) or the capacities of the nodes (higher $\alpha$). On the other hand, for the `yahoo-answers` datasets, using the same $\epsilon = 1$, the violations are practically zero for any combination of the other parameters. One reason for the difference between the violations in these two datasets may be the capacity distributions, as shown in Figure 7 in the Appendix. For all practical purposes in our envisioned application scenarios these violations are negligible.

*Any-time stopping.* An interesting property of GREEDY-MR is that it produces a feasible but suboptimal solution at each iteration. This allows to stop the algorithm at any time, or to query the current solution and let the algorithm continue in the background. Furthermore, GREEDYMR converges very fast to a good global solution. Figure 5 shows the value of the solution found by GREEDYMR as a function of the iteration. For the three datasets, `flickr-small`, `flickr-large`, and `yahoo-answers`, the GREEDYMR algorithm reaches 95% of its final $b$-matching value within 28.91%, 44.18%, and 29.35% of the total number of iterations required, respectively. The latter three numbers are averages over all the parameter settings we tried in each dataset.

# 7. CONCLUSIONS

We investigate the problem of social content matching and introduce two MapReduce algorithms: STACKMR and GREEDYMR. Both algorithms have provable approximation guarantees. In addition STACKMR provably requires a logarithmic number of MapReduce iterations.

We test our algorithms on two large real-world datasets. GREEDYMR is a good solution for practitioners. In our experiments it consistently finds the best $b$-matching value. GREEDYMR allows to query the solution at any time, which is a desirable property for real systems. Furthermore, it is easy to implement and reason about. Nevertheless, STACK-MR has high theoretical and practical interest because of its better running time. StackMR scales gracefully to very large datasets and offers high-quality results.

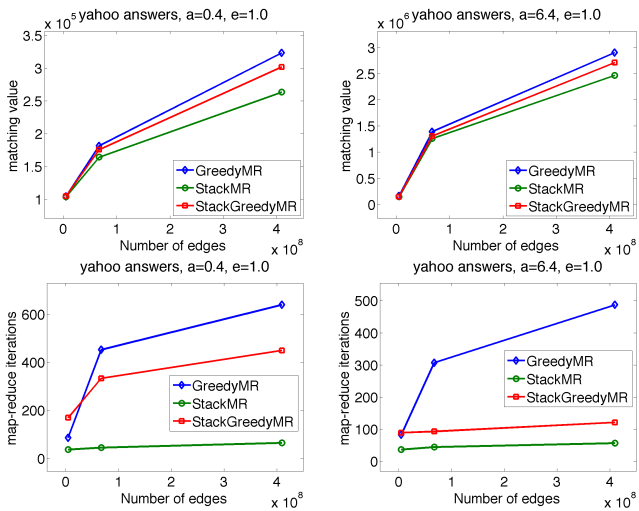

**Figure 3:** `yahoo-answers` dataset: matching value and number of iterations as a function of the number of edges.

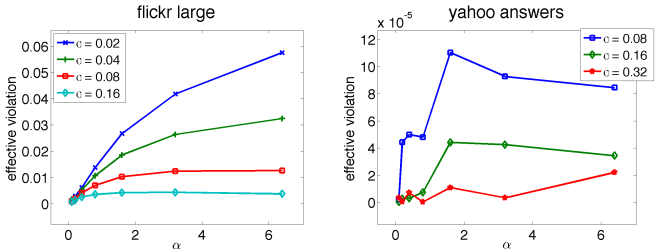

**Figure 4:** Violation of capacities for STACKMR.

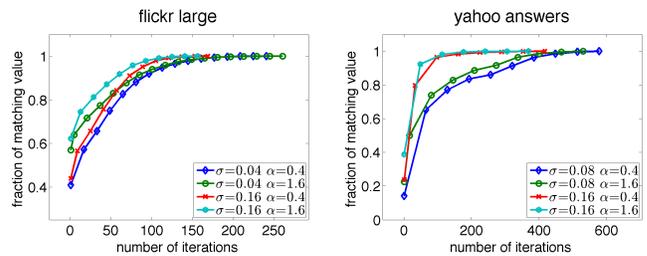

**Figure 5:** Value of the *b*-matching achieved by the GREEDYMR algorithm as a function of the number of MapReduce iterations. *b*-matching value is shown as a fraction of the final value achieved.

## 8. ACKNOWLEDGEMENTS

Gianmarco De Francisci Morales was partially supported by the EU-FP7-250527 (Assets) and the POR-FESR 2007-2013 (VISITO Tuscany) projects.

Aristides Gionis was partially supported by the Spanish Centre for the Development of Industrial Technology under the CENIT program, project CEN-20101037, "Social Media" (http://www.cenitsocialmedia.es/).

# APPENDIX

## A. ANALYSIS OF THE GREEDY ALGORITHM

In the following theorem we prove the approximation guarantee of greedy. We believe this result to be well-known, however we could not find a reference. Thus, we give a proof for completeness and self containment.

**Theorem 2** *The greedy algorithm produces a solution with approximation guarantee $\frac{1}{2}$ for the weighted b-matching problem.*

PROOF. Let $O$ be an optimum solution for a given problem instance and let $A$ be the solution yielded by the greedy algorithm. For every node $v$, let $O_v$ and $A_v$ denote the sets of edges $O \cap \Delta_G(v)$ and $A \cap \Delta_G(v)$, respectively, where $\Delta_G(v)$ is the set of edges in $G$ incident to $v$. The total weight of a set of edges $T$ is denoted by $w(T)$. We say that a node is *saturated* if exactly $b(v)$ edges of $v$ are in the greedy solution $A$ and we let $S$ denote the set of saturated nodes.

For every node $v$, we consider the sets $\widehat{O}_v \subseteq O_v \setminus A$, defined as follows: each edge $e(u,v) \in O \setminus A$ is assigned to a set $\widehat{O}_v$, for which $v$ is a saturated node and the weight of any edge in $A_v$ is larger than $w(e)$. Ties are broken arbitrarily. There must be such a node $v$, for otherwise $e$ would be included in $A$. The idea of these sets is to relate the weight of edge $e$ with the weights of the edges of $A_v$, which prevent $e$ from entering the solution. From the definition of the $\widehat{O}_v$'s it follows that

$$w(O \setminus A) = \sum_{v \in S} w(\widehat{O}_v). \qquad (6)$$

For every saturated node $v$ we have that $|O_v| \le b(v) = |A_v|$. From this and from the definition of the $\widehat{O}_v$'s we have that

$$\sum_{v \in S} w(A_v \setminus O) \ge \sum_{v \in S} w(\widehat{O}_v). \qquad (7)$$

From Equations (6) and (7) we obtain

$$
\begin{aligned}
2w(A) &\ge w(A \cap O) + \sum_{v \in S} w(A_v \setminus O) \\
&\ge w(A \cap O) + \sum_{v \in S} w(\widehat{O}_v) \\
&\ge w(O),
\end{aligned}
$$

which concludes the proof. □

The analysis is tight as proved by the following simple example. Consider a cycle consisting of three nodes $u, v, z$ and three edges $uv, vz, zu$. Let $b(u) = b(z) = 1$ and let $b(v) = 2$. Moreover, let $w(uv) = w(vz) = 1$ while $w(zu) = (1+\epsilon)$ where $\epsilon > 0$. It is easy to see that the greedy algorithm would select the edge whose weight is $1+\epsilon$, while the weight of the optimum solution is 2.

## B. ADDITIONAL PLOTS

Figure 6 shows the distribution of edge similarities for our three datasets, and Figure 7 shows the distribution of capacities.

## C. PSEUDOCODE OF OUR ALGORITHMS

Algorithms 1 and 2 present the variants of the STACKMR algorithm with no capacity violations and $(1 + \epsilon)$ capacity violations, respectively. Algorithm 3 shows the GREEDYMR algorithm.

---

**Algorithm 1** STACKMR satisfying all capacity constraints

---
1: /* Pushing Stage */
2: **while** $E$ is non empty **do**
3:   Compute a maximal $\lceil \epsilon b \rceil$-matching $M$ (each vertex $v$ has capacity $\lceil \epsilon b(v) \rceil$), using the procedure in [12];
4:   Push all edges of $M$ on the distributed stack ($M$ becomes a layer of the stack).
5:   **for** all $e \in M$ in parallel **do**
6:     Let $\delta(e) = (w(e) - y_u/b(u) - y_v/b(v))/2$;
7:     increase $y_u$ and $y_v$ by $\delta(e)$;
8:   **end for**
9:   Update $E$ by eliminating all edges that have become weakly covered
10: **end while**
11: /* Popping Stage */
12: **while** the distributed stack is nonempty **do**
13:   Pop a layer $M$ out of the distributed stack.
14:   In parallel tentatively include all edges of $M$ in the solution.
15:   If there is a vertex $v$ whose capacity is exceeded then mark all edges in $M$ incident to $v$ as *overflow*, remove them from the solution and remove all edges in $E \setminus M$ incident to $v$ from the graph.
16:   For each vertex $v$: let $\bar{b}(v)$ be the number of edges in $M$ that are incident to $v$; update $b(v) \leftarrow b(v) - \bar{b}(v)$; if $b(v) = 0$ then remove all edges incident to $v$.
17: **end while**
18: /* Computing a feasible solution */
19: **while** there are *overflow* edges **do**
20:   Let $\bar{\mathcal{L}}$ be the set of overflow edges such that for every $e \in \bar{\mathcal{L}}$ there is no overflow edge $f$, incompatible with $e$, for which $\delta(f) > (1 + \epsilon)\delta(e)$.
21:   Compute a maximal $b$-matching $\bar{M}$ including only edges of $\bar{\mathcal{L}}$ ($\bar{M}$ shall be referred to as a sublayer of the stack);
22:   In parallel, for each edge $e \in \bar{M}$, include $e$ in the solution if this maintains the solution feasible.
23:   For each vertex $v$: let $\bar{b}(v)$ be the set of edges in $\bar{M}$ that are incident to $v$ (these edges are included in the solution); update $b(v) \leftarrow b(v) - \bar{b}(v)$; if $b(v) \le 0$ then remove from the set of overflow edges all edges incident to $v$.
24:   Remove all edges in $\bar{M}$ from the set of overflow edges.
25: **end while**

---

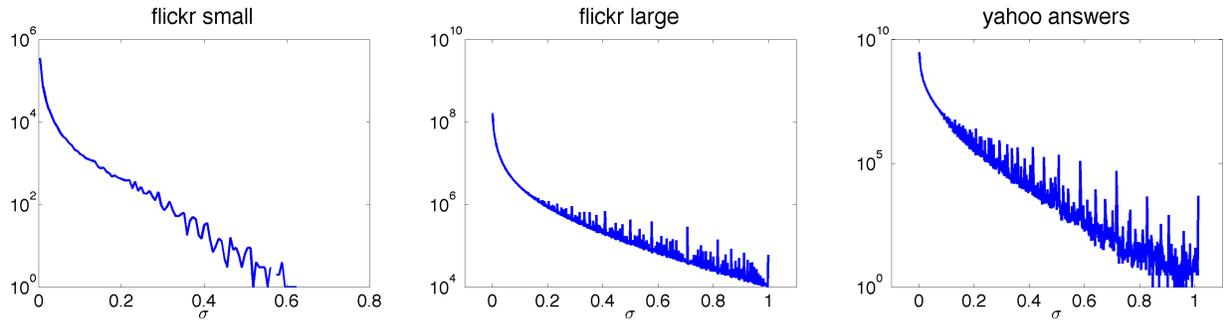

Figure 6: The distribution of edge similarities for the three datasets.

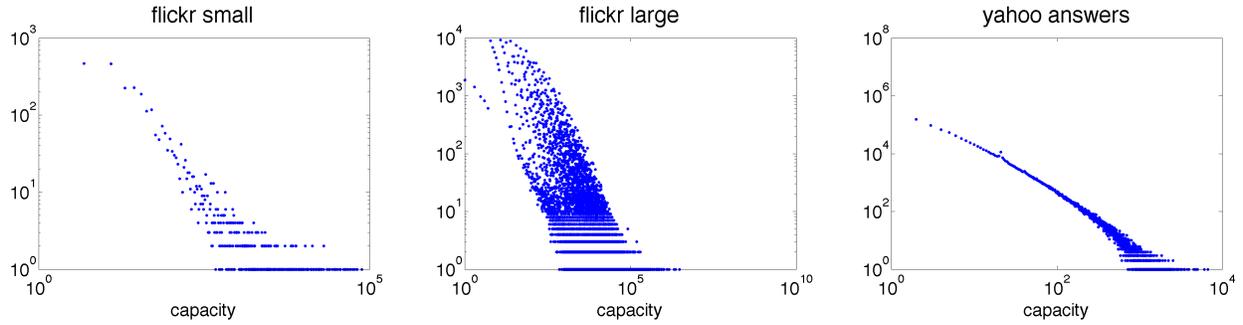

Figure 7: The distribution of capacities for the three datasets.

---

**Algorithm 2** STACKMR violating capacity constraints by a factor of at most $(1+\epsilon)$

1: /* Pushing Stage */
2: **while** $E$ is non empty **do**
3:　Compute a maximal $\lceil \epsilon b \rceil$-matching $M$ (each vertex $v$ has capacity $\lceil \epsilon b(v) \rceil$), using the procedure in [12];
4:　Push all edges of $M$ on the distributed stack ($M$ becomes a layer of the stack);
5:　**for** all $e \in M$ in parallel **do**
6:　　Let $\delta(e) = (w(e) - y_u/b(u) - y_v/b(v))\,/2$;
7:　　increase $y_u$ and $y_v$ by $\delta(e)$;
8:　**end for**
9:　Update $E$ by eliminating all edges that have become weakly covered
10: **end while**
11: /* Popping Stage */
12: **while** the distributed stack is nonempty **do**
13:　Pop a layer $M$ out of the distributed stack.
14:　In parallel include all edges of $M$ in the solution.
15:　For each vertex $v$: let $\bar{b}(v)$ be the set of edges in $M$ that are incident to $v$; update $b(v) \leftarrow b(v) - \bar{b}(v)$; if $b(v) \leq 0$ then remove all edges incident to $v$.
16: **end while**

---

**Algorithm 3** GREEDYMR

1: **while** $E$ is non empty **do**
2:　**for** all $v \in V$ in parallel **do**
3:　　Let $\widehat{L_v}$ be the set of $b(v)$ edges incident to $v$ with maximum weight;
4:　　Let $F$ be $\widehat{L_v} \cap \widehat{L_U}$ where $U = \{u \in V : \exists\, e(v, u) \in E\}$ is the set f vertexes sharing an edge with $v$;
5:　　Update $M \leftarrow M \cup F$;
6:　　Update $E \leftarrow E \setminus F$;
7:　　Update $b(v) \leftarrow b(v) - b_F(v)$;
8:　　If $b(v) = 0$ remove $v$ from $V$ and remove all edges incident to $v$ from $E$;
9:　**end for**
10: **end while**
11: return $M$;